%% file: understanding_emis_paper.tex
\documentclass[useAMS,usegraphicx,usenatbib]{mn2e}
\usepackage{graphicx}
\usepackage{amsmath}
\usepackage{amssymb}
\usepackage{subfigure}
\usepackage{url}
\usepackage{multirow}
\input{defn}

\title[Understanding X-ray reflection emissivity profiles]{Understanding X-ray reflection emissivity profiles in AGN: Locating the X-ray source}
\author[D. R. Wilkins \& A. C. Fabian]{D. R. Wilkins
  \thanks{E-mail: drw@ast.cam.ac.uk} 
and A. C. Fabian\\Institute of Astronomy, University of Cambridge, Madingley Road, Cambridge CB3 0HA}
\begin{document}

\date{Accepted 2012 May 14. Received 2012 May 14; in original form 2011 December 21}

\pagerange{\pageref{firstpage}--\pageref{lastpage}} \pubyear{2012}

\maketitle

\label{firstpage}

\begin{abstract}
The illumination pattern (or emissivity profile) of the accretion disc due to the reflection of X-rays in AGN can be understood in terms of relativistic effects on the rays propagating from a source in a corona surrounding the central black hole, both on their trajectories and on the accretion disc itself. Theoretical emissivity profiles due to isotropic point sources as well as simple extended geometries are computed in general relativistic ray tracing simulations performed on graphics processing units (GPUs). Such simulations assuming only general relativity naturally explain the accretion disc emissivity profiles determined from relativistically broadened emission lines which fall off steeply (with power law indices of between 6 and 8) over the inner regions of the disc, then flattening off to almost a constant before tending to a constant power law of index 3 over the outer disc. Simulations for a variety of source locations, extents and geometries show how the emissivity profiles depend on these properties, and when combined with reverberation time lags allow the location and extent of the primary X-ray source to be constrained. Comparing the emissivity profile determined from the broadened iron K emission line in spectra of 1H\,0707-495 obtained in January 2008 to theoretical emissivity profiles and applying constraints from reverberation lags suggest that there exists an extended region of primary X-ray emission located as low as 2\rg\ above the accretion disc, extending outwards to a radius of around 30\rg.
\end{abstract}


\section{Introduction}
When the accretion disc surrounding the central black hole of an AGN (or for that matter a galactic black hole) is illuminated by a non-thermal X-ray source in a corona of hot particles surrounding the black hole, in addition to the continuum spectrum from this source, a reflection spectrum arising from the accretion disc will be observed \citep{george_fabian}. The reflection spectrum will contain a number of emission lines and absorption features \citep{ross+99}, broadened by general relativistic effects in the proximity of the black hole and relativistic Doppler shifting due to the motion of the reflecting material in its orbit in the accretion disc \citep{fabian+89,laor-91}, the most prominent of which is the K$\alpha$ emission line of iron at 6.4\keV\ in the rest frame of the emitting material \citep{matt+97}.

In order to understand the emergent reflection spectrum, it is necessary to understand the illumination pattern of the accretion disc, that is its \textit{emissivity profile}, the reflected power per unit area as a function of location on the disc. Following \citet{laor-91}, the emissivity, $\epsilon(r)$, is defined as the radial weighting of the reflected radiation from the accretion disc, measured in the rest frame of the disc material, such that the observed spectrum is given by
\begin{equation}
	F_0(\nu_0) = \int \epsilon(r_e) I_r\left(\frac{\nu_e}{g}\right) T(r_e, g) dg r_e dr_e
\end{equation}
where $I_r$ is the rest-frame reflection emerging from the accretion disc, folded through the transfer funtion, $T(r_e,g)$ which projects the rays around the black hole to the observer, integrating over all redshifts $g$ that will shift the emitted photons into the observed energy band.

In the simplest case of a point source in flat, Euclidean spacetime, the emissivity at a point on the disc is proportional to the inverse-square of the distance from the source, multiplied by the cosine of the angle at which the ray hits the disc from the normal, giving a form $r^{-3}$ at large radius out from the source. In general relativity, however, rays will be focussed towards the black hole (and the inner disc) so naively one would expect a steeper fall-off in emissivity with distance from the black hole.

Initial models for the spectrum of relativistically broadened emission lines such as that of \citet{laor-91} assume that the emissivity profile of the accretion disc follows a power law of constant index. The classical case would be an index of $3$, though the best fits to observed spectra may require a steeper index. \citet{dovciak+04} present an extended framework for fitting the spectral profiles of relativistically broadened emission lines with a suite of models including \textsc{kyline} which can either assume an emissivity profile taking form of a power law or take a tabulated radial emissivity profile for the accretion disc. 

To reproduce the enhancement in emissivity on the inner regions of the disc due to relativistic effects, in order to correctly model the observed iron K$\alpha$ emission line in the Seyfert 1 galaxy MCG--6-30-15, \citet{fabian+02} found that a once broken power law form was required for the emissivity profile in the \textsc{laor2} broadened emission line model, with the emissivity profile falling off more steeply over the inner regions of the disc before tending to approximately the classically-expected index of around $3$ over the outer part of the disc. This approach was also taken by \citet{brenneman_reynolds} using the observed profile of the iron K$\alpha$ spectral line to measure the spin of the black hole (the \textsc{kerrdisk} model) and in the \textsc{relline} model of \citet{dauser+10}.

While the emissivity profile is typically approximated as a power law when fitting models to the reflection spectrum, the details will depend upon the location, spatial extent and geometry of the source. A number of properties of accreting black holes are determined through the detection of emission from innermost regions of the accretion disc, not least the innermost extent of the disc which is used to infer the spin of the central black hole, assuming the accretion disc extends down to the innermost stable orbit \citep{reynolds_fabian08} or to infer truncation of the accretion disc before this point. It is therefore important to understand how the emissivity profile of the disc varies over this region such that detections of emission from these radii can be correctly interpreted. For example, if too steep a falling emissivity profile is assumed, in order to correctly reproduce the flux levels emitted from the innermost regions, the disc must be truncated at a larger radius, in turn causing the spin of the black hole to be underestimated.

Previous calculations of the illumination of black hole accretion discs by a point source on the rotation axis above the black hole, \textit{e.g.} \citet{miniutti+03} and \citet{martocchia+00}, predict emissivity profiles approximating twice-broken power laws, falling steeply over the inner regions of the disc, then flattening before tending to a constant power law index over the outermost radii, while \citet{suebsuwong+06} extended this work to orbiting point sources offset from the rotation axis, yielding qualitatively similar results.

The emissivity profile of the accretion disc in the narrow line Seyfert 1 galaxy, 1H\,0707-495 was determined by independently fitting the contributions to the relativistically broadened emission lines from successive radii in the disc \citep{wilkins_fabian_2011a}. The accretion disc was found to have a steeply falling emissivity profile (with a power law index 7.8) in the inner region of the disc then flattening off to almost constant emissivity between radii of 5.6\rg\ and 34.5\rg\ before tending to a constant power law index of 3.3 (slightly steeper than the classical case) over the outermost parts of the disc, in broad agreement with previously predicted forms.

We develop here a formalism in which theoretical emissivity profiles due to point sources and extended sources located in the corona may be predicted through general relativistic ray tracing simulations. While previous theoretical work derives emissivity profiles of accretion discs for selected cases, mostly for the case of a point sources on the rotation axis above the central black hole, we systematically calculate emissivity profiles for varying locations of the primary X-ray source and further this, to include extended sources of different sizes and geometries. This allows the effects of these parameters on the emissivity profile to be studied, such that properties of the X-ray source may be inferred from the observed emissivity profile.

\section{Calculating Theoretical Emissivity Profiles}
Theoretical emissivity profiles due to the reflection of the primary X-ray continuum, from a source in a corona in the central regions of the AGN, off of an accretion disc in the equatorial plane are calculated in ray tracing simulations in the Kerr spacetime around the black hole.

\subsection{Ray tracing}
In Boyer-Lindquist co-ordinates, the Kerr metric \citep{kerr} for the spacetime around a spinning black hole (spin parameter $a = \frac{J}{M}$) is written
\begin{eqnarray*}
	ds^2 &=& \left(1-\frac{2r}{\rho^2}\right)dt^2 + \frac{4ar\sin^2\theta}{\rho^2}dtd\varphi - \frac{\rho^2}{\Delta}dr^2 - \rho^2 d\theta^2 \\
	&-& \left(r^2 + a^2 + \frac{2a^2 r \sin^2\theta}{\rho^2}\right)\sin^2\theta d\varphi^2
\end{eqnarray*}
where
\begin{eqnarray*}
	\rho^2 &\equiv& r^2 + a^2\cos^2\theta \\
	\Delta &\equiv& r^2 - 2r + a^2
\end{eqnarray*}
Taking $c=1$ and $\frac{GM}{c^2} = 1$ to work in the natural units of gravitational radii, $r_g = \frac{GM}{c^2}$, for both spacelike and timelike co-ordinates.

In this spacetime, photons propagate along null geodesics satisfying the condition
\begin{equation}
	\label{null.equ}	
	g_{\mu\nu} \dot{x}^\mu \dot{x}^\nu = 0
\end{equation}
and the geodesic equation
\begin{equation}
	\label{geodesic.equ}
	\ddot{x}^a + \Gamma^a_{\ bc} \dot{x}^b \dot{x}^c = 0
\end{equation}
where $g_{\mu\nu}$ is the metric tensor defined such that $ds^2 = g_{\mu\nu}\dot{x}^\mu\dot{x}^\nu$, and $\Gamma^a_{\ bc}$ are the connection co-efficients, defining the derivatives of the basis vectors with $\frac{d\mathbf{e}_a}{dx^b} \equiv \Gamma^c_{\ ab}\mathbf{e}_c$. Repeated upper and and lower indices imply summation.

Solving Equation \ref{geodesic.equ} with constraint \ref{null.equ}, the geodesic equations that characterise the propagation of photons can be written $\mathbf{\dot{r}}=(\dot{t},\dot{r},\dot{\theta},\dot{\varphi})$
\begin{eqnarray*}
	\label{tdot.equ}	
	\dot{t} &=& \frac{
		\left[(r^2 + a^2\cos^2\theta)(r^2+a^2) + 2a^2 r\sin^2\theta\right]k - 2arh
	}
	{ 
		r^2\left(1+\frac{a^2\cos^2\theta}{r^2} - \frac{2}{r}\right)\left(r^2+a^2\right) + 2a^2r\sin^2\theta
	} \\
	\label{phidot.equ}
	\dot{\varphi} &=& \frac{
		2ar k \sin^2\theta + (r^2+a^2\cos^2\theta - 2r)h
	}
	{
		(r^2+a^2)(r^2+a^2\cos^2\theta-2r)\sin^2\theta + 2a^2 r\sin^4\theta
	}
\\
	\label{thetadot.equ}
	\dot{\theta^2} &=& \frac{Q + (ka\cos\theta - h\cot\theta)(ka\cos\theta + h\cot\theta)}{\rho^4}
\\
	\label{rdot.equ}
	\dot{r}^2 &=& \frac{\Delta}{\rho^2}\left[ k\dot{t} - h\dot{\varphi} - \rho^2\dot\theta^2  \right]
\end{eqnarray*}
The derivatives are with respect to some affine parameter, $\sigma$ and $k$, $h$ and $Q$ are conserved constants of the motion which distinguish the geodesics \citep{carter-68}.

Given the starting point of a photon and the values of the constants of motion, a ray can be traced as the affine parameter advances by integration of the geodesic equations.
\begin{equation}
	\mathbf{r}(\sigma + d\sigma) = \mathbf{r}(\sigma) + \mathbf{\dot{r}}(\sigma) d\sigma
\end{equation}

A number of singularities exist in the Kerr spacetime when expressed in Boyer-Lindquist co-ordinates. In addition to the co-ordinate singularity defining the event horizon, from which rays cannot return once they have crossed at $r_H = 1 - \sqrt{(1-a)(1+a)}$ for spin parameter $a$, further co-ordinate singularities exist at $\theta = 0$ and at $\varphi = 0,2\pi$ (where the co-ordinate is discontinuous). It is necessary to integrate the geodesic equation with greater numerical precision closer to the singularities, to prevent the solution diverging here, but for economy in the computation, a larger step size can be used further from these singularities. As such, a variable step size in the affine parameter is used, proportional to the distance from the singularity divided by the velocity in that co-ordinate, given a baseline precision parameter $\tau$, defined such that
\begin{equation}
	d\sigma = \frac{\left| \frac{r - r_H}{\dot{r}} \right|}{\tau}
\end{equation}
The appropriate step size is computed in this manner for each of the $r$, $\theta$ and $\varphi$ co-ordinates, and the step size is taken to be whichever of these is the smallest.

\subsection{The X-ray source}
An isotropic point source is defined as emitting equal power into equal solid angle in its own instantaneous rest frame, where the solid angle element for polar angles $\alpha$ and $\beta$ is
\[ d\Omega' = d(\cos\alpha)d\beta \]
The source frame is defined by a tetrad of orthogonal basis vectors. The timelike tetrad basis vector, $\mathbf{e}'_{(t)}$ is, by definition of the instantaneous rest frame, parallel to the source's 4-velocity, $\mathbf{u}_0$ (since in its rest frame, the 4-velocity must have no spacelike component as its 3-velocity is zero). By the equivalence principle which states that in a local, freely-falling laboratory, the laws of physics reduce to those of special relativity, the local spacetime in the source frame is flat and the tetrad (unit) basis vectors satisfy
\begin{equation}
	\mathbf{e}'_{(a)} \cdot \mathbf{e}'_{(b)} = g_{\mu\nu} e_{(a)}^\mu e_{(b)}^\nu = \eta_{(a)(b)}
	\label{orthcondition.equ}	
\end{equation}
Note that bracketed subscripts indicate directions in the tetrad basis and those not bracketed indicate directions in the co-ordinate basis. $\eta_{(a)(b)} = \mathrm{diag}(1,-1,-1,-1)$ is the Minkowski (flat) space metric. The $z$-axis of the source frame is constructed parallel to the radial basis vector in the (Boyer-Lindquist) co-ordinate frame (see Appendix).

In the source frame, rays with photon energy $E$ are started at equal intervals in solid angle, that is equal intervals in $\cos\alpha$ and $\beta$, with 4-momentum
\begin{equation}
	\mathbf{p} = \left( E, E\sin\alpha\sin\beta, E\sin\alpha\cos\beta, E\cos\alpha \right)
\end{equation}

The ray is then transformed back to the global co-ordinate basis and given the source location, the constants of motion for the ray can be found by inverting the equations of motion, allowing the ray to be propagated.

\subsection{Emissivity Profile Calculation}
Photons are propagated until they reach the disc, when $\theta\ge\frac{\pi}{2}$, or until they escape to a maximum allowed radius (or the maximum number of allowed steps is reached, to prevent the code entering an infinite loop if the step size were to become very small). When the photon hits the disc, its position is assigned to a radial bin and the number of photons hitting the disc in each bin is counted.

Once the number of photons in each radial bin has been counted, the emissivity profile is obtained by dividing these counts by the area of each bin (an annulus). Strictly speaking, that is the proper area of the annulus, as measured in the disc frame, from the definition of the emissivity profile. The proper area as measured by a stationary observer is obtained from the metric. For radial bins of the accretion disc, $dt = d\theta = 0$, so the area element is
\[ d^2x = \sqrt{g_{rr}g_{\varphi\varphi}}drd\varphi \]
\begin{equation}
	d^2x = \frac{\rho}{\sqrt{\Delta}}\sqrt{r^2 + a^2 + \frac{2 a^2 r}{\rho^2}}drd\varphi
\end{equation}
So the area of an annulus of co-ordinate thickness $dr$ at radial co-ordinate $r$ is 
\begin{equation}
	A(r,dr) = 2\pi\frac{\rho}{\sqrt{\Delta}}\sqrt{r^2 + a^2 + \frac{2 a^2 r}{\rho^2}}dr
\end{equation}

In addition to the warping of spacetime around the black hole, the proper area of an orbiting element of the disc will be Lorentz contracted according to a stationary observer, so in the disc frame, the area will be increased by a factor of $\gamma$, the Lorentz factor of the orbiting disc element.

A further subtlety lies here, and that is the velocity of the `stationary' observer mentioned above. In the Kerr spacetime, the rotation of the black hole causes `frame-dragging' of an observer whose angular momentum is zero. Therefore it is more meaningful to define the stationary observer as one who is dragged in this way, rotating at $\frac{d\varphi}{dt} = \omega = \frac{2ar}{\Sigma^2}$ (where $\Sigma^2 \equiv \left(r^2 + a^2\right)^2 - a^2\Delta\sin^2\theta$). In fact, the solution $\frac{d\varphi}{dt} = 0$ is not possible within the stationary limit surface upon which $g_{tt}=0$ (this lies at 2\rg\ for a maximally rotating Kerr black hole) --- massive particles are forced to orbit the black hole within this region since there is no allowed solution of the geodesic equations which have zero 3-velocity.

The Lorentz factor of the orbiting disc element is calculated from the disc velocity with respect to an observer with $\frac{d\varphi}{dt} = \omega$. Solving the geodesic equations for a massive particle in a circular orbit ($\dot{r}=0$) in the equatorial plane, the 4-velocity of the disc element is
\[ \mathbf{v}_d = (\dot{t},0,0,\dot{\varphi}) = \dot{t}(1,0,0,\Omega) \] 
\[ \Omega = \frac{d\varphi}{dt} = \left(a \pm r^\frac{3}{2}\right)^{-1} \]
(with the $+$ and $-$ signs in the denominator corresponding to prograde and retrograde orbits respectively). This is then projected into the frame of the observer (whose tetrad basis is constructed as for the X-ray source) by taking the appropriate scalar products.
\[ v^{(a)} = \mathbf{v}\cdot\mathbf{e}'_{(a)} = g_{\mu\nu}v^\mu e_{(a)}^\nu \]
The Lorentz factor is then calculated from the 3-velocity, $\overrightarrow{v} = (v^{(1)}, v^{(2)}, v^{(3)})$ and its squared magnitude, $v^2 = (v^{(1)})^2 + (v^{(2)})^2 + (v^{(3)})^2$.
\[ \gamma = \frac{1}{\sqrt{1 - \frac{v^2}{c^2}}} \]

In addition to the relativistic effects on the areas of the radial bins, the energy of individual rays will be red or blueshifted as rays travel further from or closer to the black hole due to the variation in the rates at which the proper times of observers elapse. Since the emissivity is defined as the flux emitted from the disc (proportional to the flux received for reflection) and the flux is the power per unit area, it is proportional to the product of the photon arrival rate and the energy of each photon. The emissivity is enhanced by a factor of $g^{-2}$, where $g\equiv\frac{\nu_\mathrm{E}}{\nu_\mathrm{O}}$ is the ratio of the emitted and observed photon energies. The first factor of $g$ arises from the shifts in the energy of individual photons, while the second is due to the photon arrival rate along each ray. The redshift is calculated by the projection of the photon 4-momentum on its geodesic on to the observers' timelike axes since the photon energy is the $p^0$ component of the 4-momentum.
\begin{equation}
	\label{redshift.equ}
 g^{-1} \equiv \frac{\nu_\mathrm{O}}{\nu_\mathrm{E}} = 
 \frac{\mathbf{v}_\mathrm{O}\cdot\mathbf{p}(\mathrm{O})}{\mathbf{v}_\mathrm{E}\cdot\mathbf{p}(\mathrm{E})} = \frac{g_{\mu\nu}v_O^\mu p^\nu(O)}{g_{\rho\sigma}v_E^\rho p^\sigma(E)}
\end{equation}
where $\mathbf{v}_\mathrm{O}$ and $\mathbf{v}_\mathrm{E}$ are, respectively, the 4-velocities of the observer (the disc element) and emitter (the primary X-ray source), while $\mathbf{p}(x)$ is the photon 4-momentum as a function of position along the geodesic, here taken at observation and emission.

Putting this together, if the photon count in a radial bin at co-ordinate radius $r$ and of thickness $dr$ is $N(r,dr)$, the emissivity profile is given by
\begin{equation}
	\label{emis_calc.equ}	
	\epsilon(r) = \frac{N(r,dr)}{g^2A(r,dr)}
\end{equation}

\subsection{Implementation}
The above algorithm is coded to run on Graphics Processing Units (GPUs) using the \textsc{nvidia cuda} programming architecture. A typical GPU at the time of writing has between 256 and 512 processing cores, each capable of running an independent computing thread, meaning that parallelised computations can often be sped up by factors of several hundred or more. In the \textsc{cuda} programming model, threads are conceptualised on a two-dimensional grid, meaning each thread can be readily mapped on to a point in the source-frame trajectory $(\cos\alpha, \beta)$ parameter space.

In order to compute emissivity profiles due to an X-ray source from a given set of rays, each ray is traced by an independent computing thread on the GPU. The parameters (starting co-ordinates and constant of motion) are computed by the conventional CPU and set up in the memory on board the GPU, with memory addresses allocated to each thread (\textit{e.g.} for an isotropic point source, all the rays are started at the same spacetime location and the constants of motion are incremented according to equal steps in $\cos\alpha$ and $\beta$). The computing threads are then executed, moving each ray as the affine parameter advances according to the equations of motion, until they reach their limits on the disc, at the event horizon or at a maximum allowed radius or number of steps. Once each thread has finished, the final positions of the rays are read back from the GPU memory and binned into locations on the accretion disc allowing the emissivity profile to computed.

\section{Theoretical Emissivity Profiles}

\subsection{Relativistic Effects}
\begin{figure}
	\centering
	\includegraphics[width=85mm]{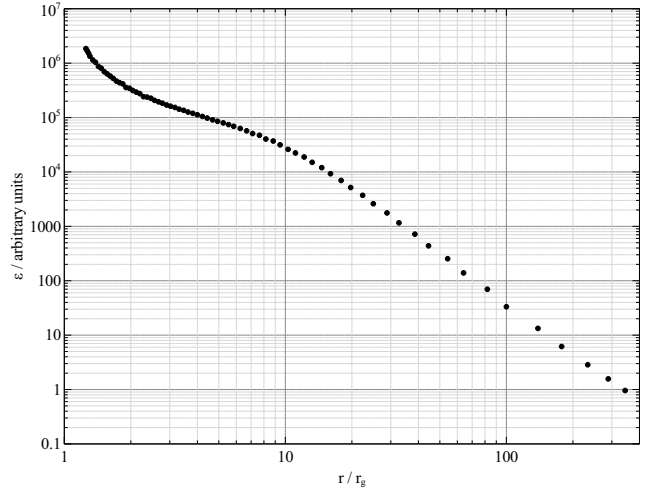}
	\caption{Theoretical accretion disc emissivity profile due to a stationary isotropic point source located at a height of 10\rg\ on the rotation axis.}
	\label{emis_ax_10.fig}
\end{figure}

Computing the emissivity profile resulting from an isotropic point source at a height of 10\rg\ on the rotation axis above the plane of the accretion disc (Fig. \ref{emis_ax_10.fig}) illustrates the key effects that influence the form of the emissivity profile that will enable observed forms to be explained in terms of the properties of the X-ray source.
\begin{figure}
	\centering
	\includegraphics[width=85mm]{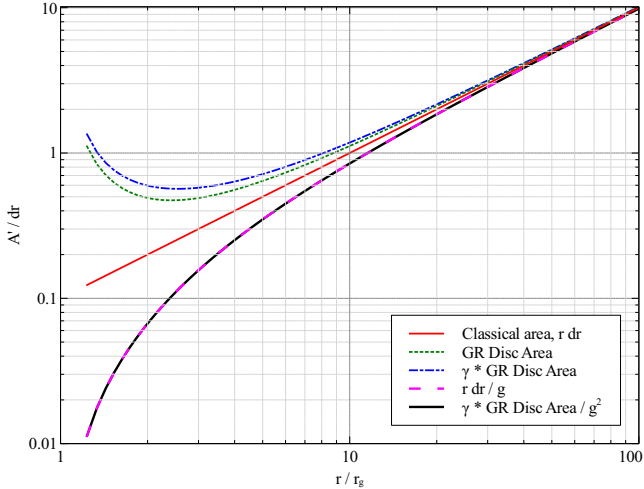}
	\caption{The effective area of successive annuli in the accretion disc, that is the total value (per $dr$) by which the photon count in the radial bin from the ray tracing simulation is divided in order to obtain the emissivity profile in Equation \ref{emis_calc.equ}. The classical area, $r dr$ of the annulus is compared with that in the Kerr spacetime for a stationary area element as well as an orbiting element, multiplying by the Lorentz factor, $\gamma$, to take into account Lorentz contraction as observed by a stationary observer. Finally, the effect of redshift on the rays is accounted for. It is noted that the relativistic effects on the area of the disc element exactly cancel one factor of the redshift, $g$, such that the fully relativistic result is equal to  $rdr/g$.}
	\label{disc_area.fig}
\end{figure}

In a flat, Euclidean spacetime the flux received from the source at each point on the disc and thus the reflected flux (emissivity) from that point will vary simply as the inverse square of the distance from the primary X-ray source, projected into the direction normal to the disc plane. If the source is at a height $h$ above $r=0$ in the disc, the emissivity profile will go as $(r^2 + h^2)^{-1}\cos\vartheta$, with $\cos\vartheta=\frac{h}{\sqrt{r^2+h^2}}$, the angle from the normal at which the ray hits the disc. This will be constant (a flat profile) in the limit $r \ll h$ (the inner disc) and tending to $r^{-3}$ at large radius, giving a power law emissivity profile with index $3$ as can be seen where the flatter region of the emissivity profile becomes a power law of constant index at the break point, $r\sim 10r_\mathrm{g}$ (approximately the height of the X-ray source above the accretion disc).

The index of the power law over the outer region of the disc is, however, slightly steeper than the classical case of 3.0, with a value of around 3.2. In the presence of the black hole, gravitational light-bending will act to focus light rays from the point source towards the black hole and will therefore increase the flux incident on the inner regions of the disc relative to the outer regions causing it to fall off faster, steepening the emissivity profile.

The contributions of the relativistic effects on the accretion disc itself are best illustrated by considering the `effective area' of the annuli in the accretion disc, that is the total value (per $dr$) by which the photon count in the radial bin from the ray tracing simulation is divided in order to obtain the emissivity profile in Equation \ref{emis_calc.equ}. Fig. \ref{disc_area.fig} shows the classical area, $r dr$ of the annulus, compared with the proper area in the Kerr spacetime for a stationary area element as well as an orbiting element, multiplying by the Lorentz factor, $\gamma$, to take into account Lorentz contraction as observed by a stationary observer.

On including the shift in photon energy and arrival rate (factor of $g^{-2}$ from Equation \ref{redshift.equ}) as rays travel towards the black hole, the cause of the steep emissivity profile over the central regions of the disc is apparent. The increase in photon energy as well as the increased photon arrival rate along each ray as measured by observers closer to the black hole with more slowly elapsing proper times, means that the flux received by the disc is greatly enhanced on the inner parts, increasing reflection from these regions (notwithstanding the effects of radiation transport out to the observer at infinity, as the emissivity profile is defined in the local rest frame of the emitting material before the transfer to the observer is included in the formation of the relativistic reflection spectrum).

It is interesting to note that the the general relativistic effects on the area of orbiting annuli in the accretion disc are cancelled exactly by one factor of redshift, so in practice the effective area of the annulus is given by dividing the classical area ($rdr$) by one factor of the redshift. This function is also plotted for comparison.

\subsection{Axial X-ray Sources}
\begin{figure}
	\centering
	\includegraphics[width=60mm]{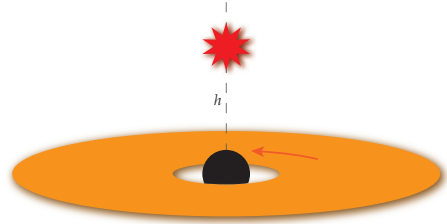}
	\caption{The geometry of an isotropic point source located on the rotation axis above the black hole for which theoretical emissivity profiles are initially computed.}
	\label{ax_source.fig}
\end{figure}

The simplest, idealised case is that of an isotropic point source, stationary upon the rotation axis (Fig. \ref{ax_source.fig}). This will allow us to explore the effects of ray propagation in the Kerr spacetime on the emissivity profile of the accretion disc, with the fewest free parameters and assumptions about the nature of the source itself. A localised source, however, may be expected if the X-ray emission results from magnetic reconnection events within the corona \citep{galeev+79,merloni_fabian}.

\begin{figure*}
\centering
\begin{minipage}{170mm}
\subfigure[]{
\includegraphics[width=85mm]{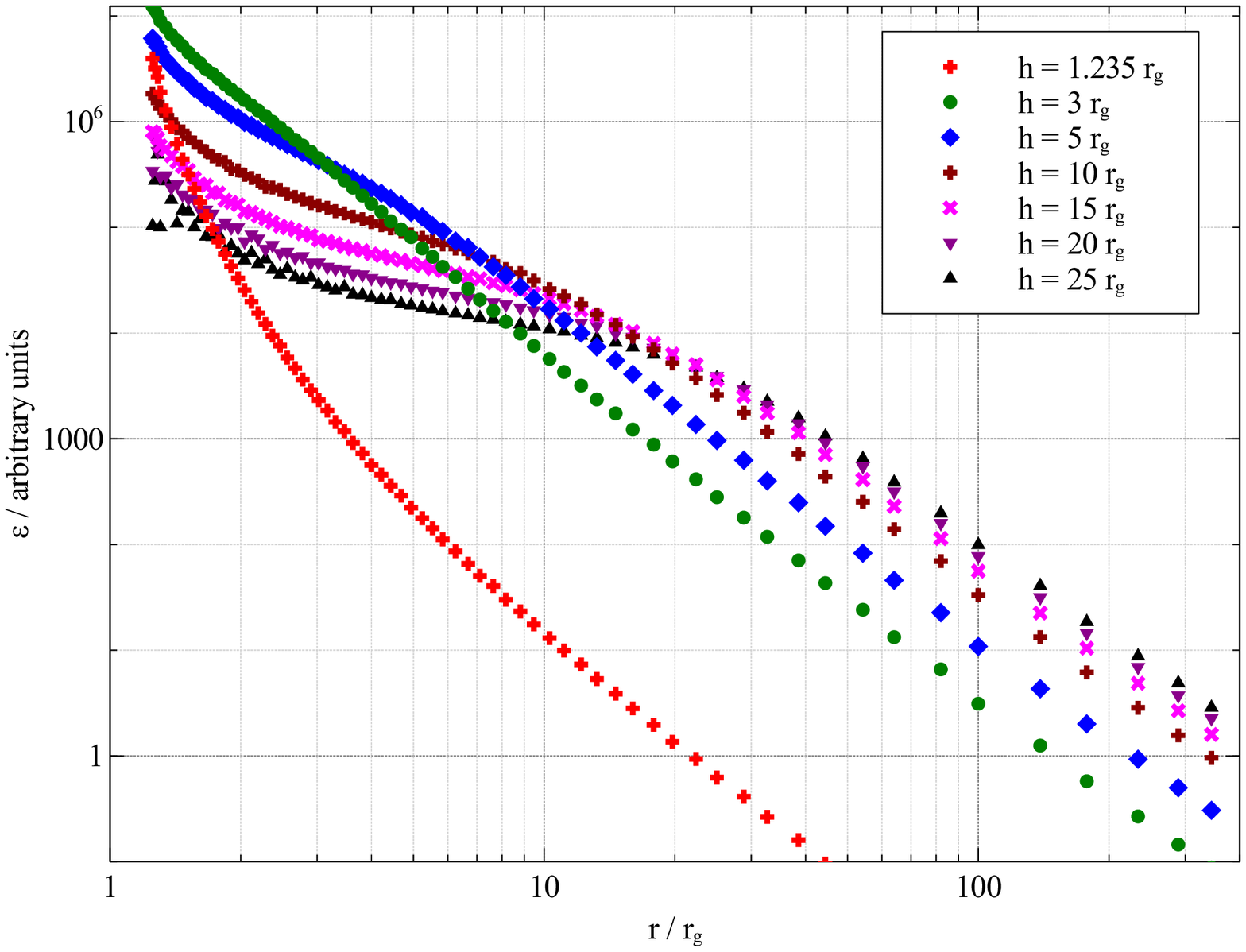}
\label{emis_ax.fig:emis}
}
\subfigure[]{
\includegraphics[width=85mm]{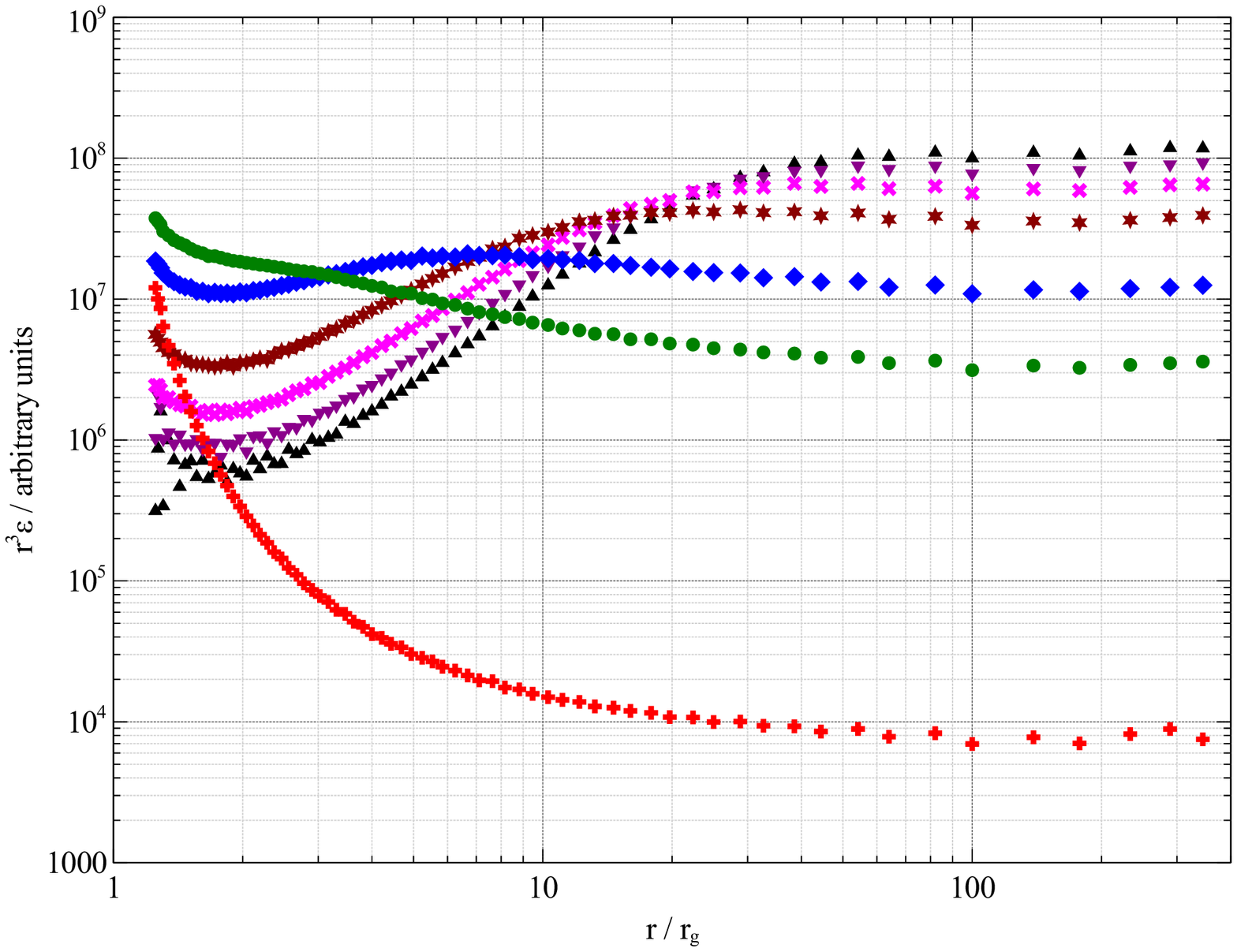}
\label{emis_ax.fig:rcube}
}
\caption[]{\subref{emis_ax.fig:emis} Theoretical accretion disc emissivity profiles due to a stationary isotropic point sources located at varying heights above the black hole on the rotation axis. \subref{emis_ax.fig:rcube} The emissivity profiles scaled by $r^3$, illustrating the steepening of the emissivity profile over the inner disc (where the plot decreases) before the profile flattens (plot increasing) and then tends to a constant power law index of around 3 over the outer disc (plot approximately constant). The plot styles in the two plots correspond to one another.}
\label{emis_ax.fig}
\end{minipage}
\end{figure*}

Theoretical emissivity profiles for isotropic point sources, stationary upon the rotation axis, at varying heights above the black hole are shown in Fig. \ref{emis_ax.fig}.

In all cases, the profile tends to a power law with an index slightly steeper than 3 over the outer regions of the disc (from 3.1 for a source at height 10\rg\ to a steeper index of 3.3 for a source height of 3\rg), with the power law indices steepening as high as $6\sim 7$ over the innermost parts.

\begin{figure}
	\centering
	\includegraphics[width=85mm]{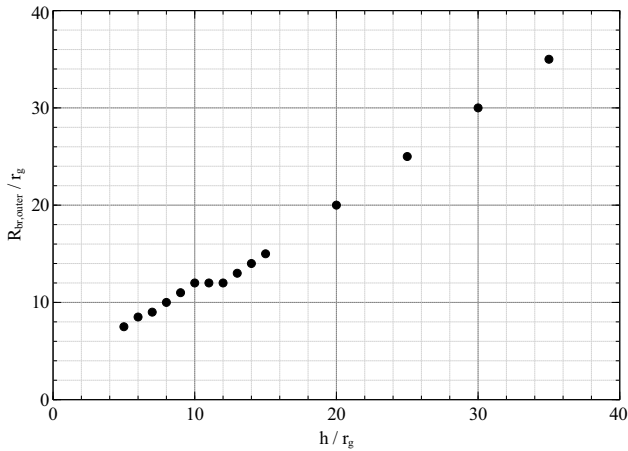}
	\caption{Observed location of the outer break point between the flattened region of the emissivity profile and the power law of constant index around 3 over the outer part of the accretion disc (where $r \gg h$) for axial point sources of increasing height above the disc.}
	\label{outer_break.fig}
\end{figure}

It can be seen that as the source is moved higher up the rotation axis, further from the black hole, the region over which the emissivity profile is flattened increases (the region where $r \ll h$). The outer break-point in the power law form of the profile moves to coincide approximately in radius with the height of the source (Fig. \ref{outer_break.fig}) for sources at heights greater than 12\rg\, with the outer break at a radius slightly greater than the height of the source for lower sources (due to the requirement for the break being $r\gg h$ which requires a greater increment in $r$ for smaller $h$).

As the source is moved closer to the black hole, the steepening over the inner regions is greatly enhanced as more photons are focussed on to the inner regions of the disc while rays are bent towards the central black hole, until the source is as low as 3\rg\ at which point the steepening is so significant that it masks the classically-predicted flattened region of the profile (now constrained to a much smaller region of the disc where $r \ll h$). The emissivity profile more closely resembles a once-broken power law which in the extreme case of a source at only 1.235\rg\ above the black hole has a very steep power law index of around 8 out as far as a radius of 5\rg\ before tending towards an index around 3.5 over the outer regions of the disc.

\subsection{Orbiting Sources}
While it is conceivable that X-rays originate from a stationary point source close to the rotation axis, one might also consider the case of a point source located elsewhere in the corona. The concept of a stationary object in the corona (if not on the rotation axis) is somewhat unphysical, given that the AGN is considered to have formed from the gravitational collapse of material towards the galactic centre, a process in which it is likely to rotate, conserving angular momentum. Furthermore, without rotation, the material will just fall into the black hole and will not survive for long as a corona (unless it is replenished). Due to the axisymmetry of the Kerr spacetime, a point source in orbit at a given radius is equivalent to a continuous ring source of that radius (which may be relevant when considering the total emission from many localised flaring events when looking at time-averaged X-ray spectra).

We again consider the idealised case of an isotropic point source to explore the observed effects in the emissivity profile due to ray propagation to the accretion disc in general relativity with the minimum number of free parameters. Orbiting point sources at various locations in the corona will be the building blocks for extended X-ray sources.

\begin{figure}
	\centering
	\includegraphics[width=60mm]{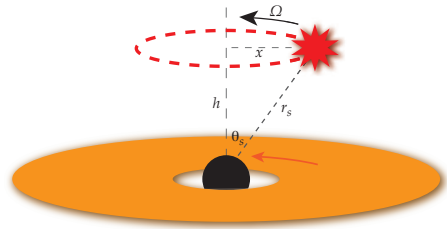}
	\caption{The considered geometry of an isotropic point source orbiting the rotation axis above the plane of the accretion disc.}
	\label{ring_source_point.fig}
\end{figure}

\begin{figure}
	\centering
	\includegraphics[width=85mm]{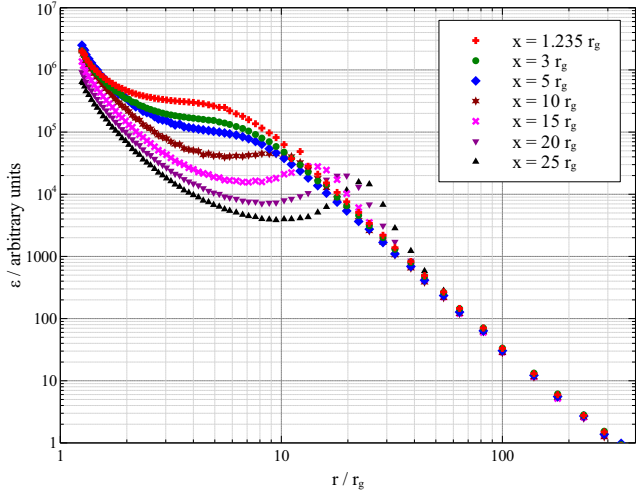}
	\caption{Theoretical emissivity profiles for isotropic point sources orbiting the rotation axis at varying radii at a height $h=5r_\mathrm{g}$ above the disc plane. The sources are `co-rotating' with the element of the accretion disc at the same radius.}
	\label{emis_h5.fig}
\end{figure}

Theoretical emissivity profiles for isotropic point sources as shown in Fig. \ref{ring_source_point.fig}, orbiting the rotation axis at varying radii at a height $h=5r_\mathrm{g}$ above the disc plane are shown in Fig. \ref{emis_h5.fig}. The sources are `co-rotating' with the element of the accretion disc at the same radius, \textit{i.e.} a source at a distance $x$ from the rotation axis as measured along a plane parallel to the accretion disc below is taken to be orbiting at the same velocity as the element of the disc in a (relativistic) Keplerian orbit at radius $x$. Where the orbiting source is close to the disc, this serves as an approximation to the orbital velocity, however if the X-ray source originates from flaring due to magnetic reconnection in poloidal field lines anchored to the ionised accretion disc, it may also be expected that the coronal material will move along with the orbiting disc.

As for the case of an axial source, the profiles are steepened from the classical case due to gravitational light bending, focussing more rays on to the inner region of the disc as well as the blueshifting of photons as they travel towards the innermost regions. These calculations show that the power law index of the emissivity profile over the innermost regions of the disc ranges from 7 for sources close to the rotation axis and black hole, within 5\rg, to 6 where the source located further out between 20 and 25\rg\ (this central steepening of the emissivity profile is almost entirely due to the blueshifting of photon energies and arrival rates as the rays travel towards the black hole now the source is located away from the inner parts of the disc). The index of the emissivity profile for the outer disc is 3.2 in each case.

The emissivity profiles are flatter over the middle region than for axial sources, owing to the source being located further out from the rotation axis, so the classical region where $h \gg r$ giving a constant flux at the disc is not masked by the relativistic effects steepening the profile over the inner disc.

Since an X-ray source close to the innermost stable orbit will be travelling at relativistic speeds if in a Keplerian orbit, emission will be `beamed' into the forward direction of source motion. This will enhance the flux in front of the source and (to a lesser extent) on the region of the disc directly below the source locus while reducing the flux received outside of this region. This will serve to further flatten the emissivity profile.

Following this reasoning, increasing the radius of the ring source flattens the profile out to a larger radius. Where the source height is less than or of the order of the source's radius, a small peak is observed in the emissivity profile below the source, where the proximate regions of the disc subtend a large solid angle at the source so a large number of rays are intercepted by this region of the disc before they are able to propagate further out.

\subsection{Extended Sources}

While it is instructive to consider the case of isotropic point sources in the corona to illustrate the factors affecting the emissivity profile of the accretion disc from X-ray reflection, in reality, the source is likely to be extended over a region of the corona where particles are heated and accelerated, up-scattering seed photons to the observed X-rays.

In the simplest case, an extended X-ray source can be considered as the sum of point sources assuming the corona is optically thin to the emitted X-rays such that emission from any part of the source can escape without interacting with other parts (as one might expect for an optically thin corona of hot, accelerated electrons that inverse-Compton scatter photons). This simple model will illustrate the behaviour of such systems in determining the accretion disc emissivity profile.

\begin{figure}
	\centering
	\includegraphics[width=85mm]{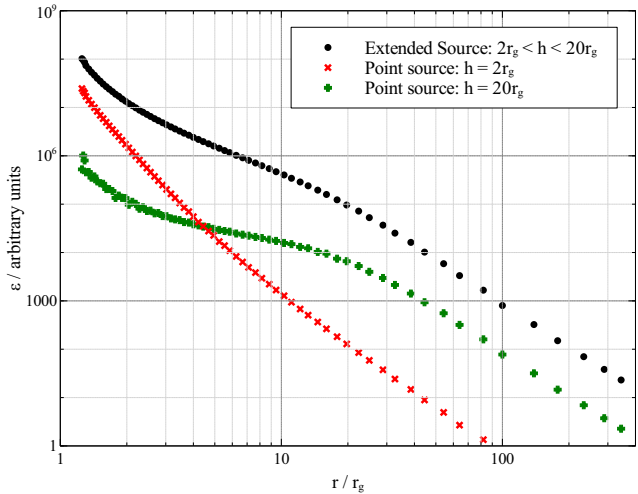}
	\caption{Theoretical accretion disc emissivity profile due to a vertically extended stationary source on the rotation axis, extending from 2\rg\ to 20\rg. The emissivity profile compared to those of point sources located at either end of the extended source, illustrating that the overall emissivity profile combines the effects of the more extreme steepening over the inner disc from the source closer to the black hole, with the slight outer break point corresponding to that in the emissivity profile of the  highest source (though the effect of summing the sources up the axis is to steepen the middle region of the profile, making the break points in the power law form less pronounced).}
	\label{emis_sum_jet.fig}
\end{figure}

A vertically extended source of X-ray emission can be constructed by summing isotropic point sources at regular intervals. This will allow the emissivity profiles of sources of finite height above the disc plane to be investigated. The emissivity profile resulting from a stationary, vertically extended source whose luminosity is constant along its length is shown in Fig. \ref{emis_sum_jet.fig}. These profiles take an almost constant power law index all the way out on the disc as the profiles from sources at successive heights sum together, masking the flattened part of each. There is steepening due to relativistic blueshift on the innermost parts and a very slight break to the outermost index of around 3 corresponding to the break point in the emissivity profile of the uppermost extent of the source.

\begin{figure}
	\centering
	\includegraphics[width=60mm]{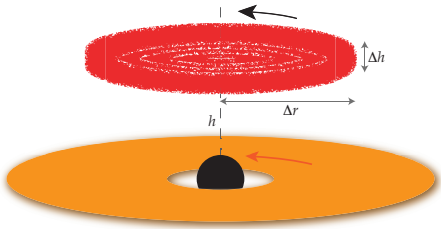}
	\caption{Extended X-ray source defined by a lower and upper height from the disc plane as well as an inner and outer radius.}
	\label{radial_extended_source.fig}
\end{figure}

Spatially extended X-ray sources may be studied in more detail through Monte-Carlo ray tracing simulations. Rays are started at random locations (with a uniform probability density function such that all locations are equally likely) within an allowed cylindrical source region (defined by a lower and upper height from the disc plane as well as an inner and outer radius) and are assigned random initial direction cosines $\cos\alpha$ and $\beta$, again with uniform probability density (Fig. \ref{radial_extended_source.fig}. This will simulate the effect of an X-ray source of finite spatial extent that is, again, optically thin to the X-rays it emits. Each local region of the source is taken to be co-rotating with the element of the accretion disc in a relativistic Keplerian orbit at the same radius. To account for the variation in source luminosity across its extent, the rays reaching the disc can be weighted by power laws in the height and radius of their origin in the source (or indeed weighted by other functions) allowing for simple non-uniform sources to be modelled, however as this introduces a number of free parameters that will not be constrained by the current quality of observations, the simplest case of constant luminosity sources is taken here to explore the effects of spatially extending the source.

\begin{figure}
	\centering
	\includegraphics[width=85mm]{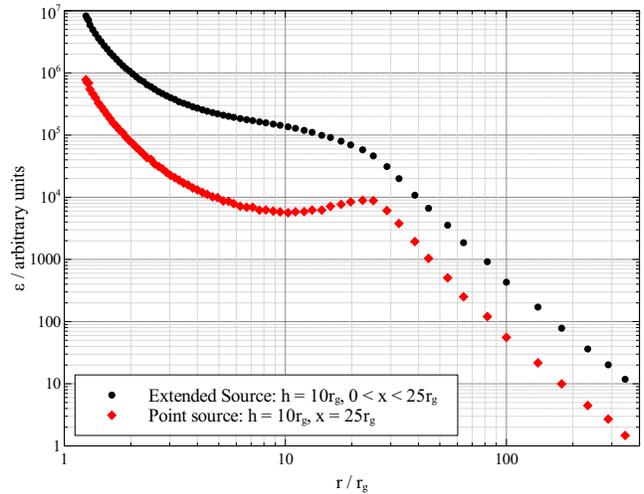}
	\caption{Theoretical accretion disc emissivity profiles due to an extended disc of emission located at a height of 10\rg\ above the disc plane and extending radially to 25\rg\ from the rotation axis, compared to that arising from a point source orbiting on the outermost edge of this source.}
	\label{emis_sum_h10r25.fig}
\end{figure}

Modelling a radially extended source at a height of 10\rg\ and extending radially to 25\rg\ produces an emissivity profile as shown in Fig. \ref{emis_sum_h10r25.fig} and, for reference, is compared to a single point source located at the outer extent. It can be seen that the outermost break point in the power law form is determined by the outer extent of the source, while the existence of X-ray sources within this radius cause the profile to be flattened off within this part before steepening, once again, over the inner regions of the disc.

\subsection{Jet Sources}
X-ray emitting jets of collimated particles accelerated to relativistic velocities are observed in a number of galactic black hole binaries such as Cygnus X-1. It is conceivable that it is the X-rays from this jet that illuminate the accretion disc giving rise to the reflection dominated component of the spectrum. A vertically collimated X-ray source could also represent, for example, the X-ray emission from a vertically collimated jet of particles accelerated up along the rotation axis of the black hole in a radio galaxy.

\begin{figure}
	\centering
	\includegraphics[width=85mm]{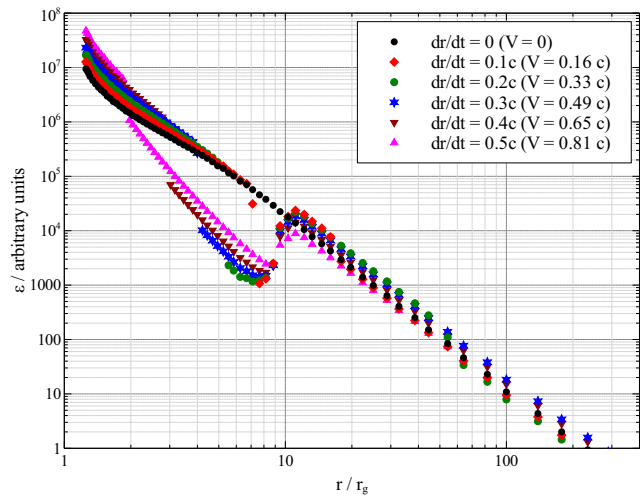}
	\caption{Theoretical accretion disc emissivity profiles due to a `jet' component moving radially up the rotation axis. The moving point source is taken to be at a height of 5\rg\ and is assumed to be isotropic in its instantaneous rest frame. Velocities are quoted in the global Boyer-Lindquist co-ordinates ($dr/dt$) as well as those measured by a stationary observer at the same location ($V$). The rays propagating backwards, towards the black hole, are still focussed onto the inner parts of the accretion disc, however as the relativistic motion of the source outwards causes emission to be beamed away from the accretion disc causing a sudden drop in the emission reaching the disc over the middle region.}
	\label{emis_jet.fig}
\end{figure}

Fig. \ref{emis_jet.fig} shows the accretion disc emissivity profiles due to illumination by point sources moving radially at a constant velocity up the rotation axis (the point sources are assumed to be isotropic in their instantaneous rest frames) compared to a stationary source. The emissivity profile are computed, as before, by constructing the tetrad basis vectors in the instantaneous rest frame of the X-ray source, now moving radially.

\begin{figure}
	\centering
	\includegraphics[width=85mm]{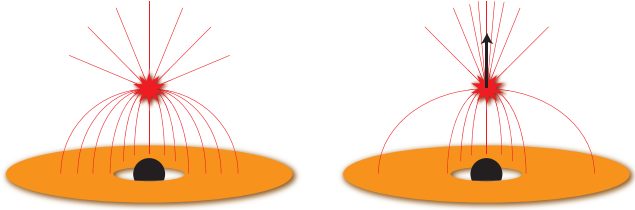}
	\caption{Relativistic motion of the jet source causes the emission to be beamed into the forward direction of motion. The black hole still focusses the backward emission onto the innermost regions of the accretion disc, however X-rays are now beamed away from the accretion disc, reducing reflection from the middle region.}
	\label{jet_source_beaming.fig}
\end{figure}

Gravitational light bending and blueshifting of rays still enhances the emission reaching the very inner part of the disc, however the highly relativistic radial motion of the sources causes the emission to be beamed in front of the motion, greatly reducing the emission behind the jet that reaches the middle region of the accretion disc (Fig. \ref{jet_source_beaming.fig}). The emission reaching the accretion disc drops dramatically by $1.5\sim 2$ orders of magnitude on the middle parts.  

The very steep fall off in the emissivity profile, the form of the profile greatly differing from observed emissivity profiles and the low fraction of the initial radiation that is reflected suggests that relativistically moving jet sources are not responsible for the significant reflection components observed in AGN and galactic black hole binaries, rather these more likely occur from a more slowly moving corona.

Accelerating jets of particles do, however, raise the possibility that the observed X-ray continuum and reflected component could be disconnected, with a fairly constant reflection component from a steady base of a jet or extended corona while the observed variable continuum radiation is dominated by fast moving particles in the jet accelerated by a variable mechanism. Weaker variation of the reflection component that is uncorrelated with the variation in the primary continuum was observed by \citet{vaughan_edelson} and \citet{fabian_vaughan-03} in MCG--6-30-15. Such disconnection of the observed continuum and reflection also relaxes the constraint on the relative fluxes observed in the continuum and reflection components of the spectrum if they are no longer dominated by the same component.

\subsection{Black Hole Spin}
The most significant effect of varying the spin of the black hole is to change the location of the innermost stable circular orbit (ISCO). For a maximally rotating black hole ($a=0.998$) in the Kerr spacetime, the ISCO lies at a radial co-ordinate of 1.235\rg\ while for a non-rotating, Schwarzschild black hole, the ISCO moves out to 6\rg. The accretion disc cannot exist stably within the ISCO since material is unable to maintain a stable orbit here and will plunge into the black hole, reducing the density in this region causing little or no reflection to be seen within the ISCO.

\begin{figure*}
\centering
\begin{minipage}{170mm}
\subfigure[]{
\includegraphics[width=80mm]{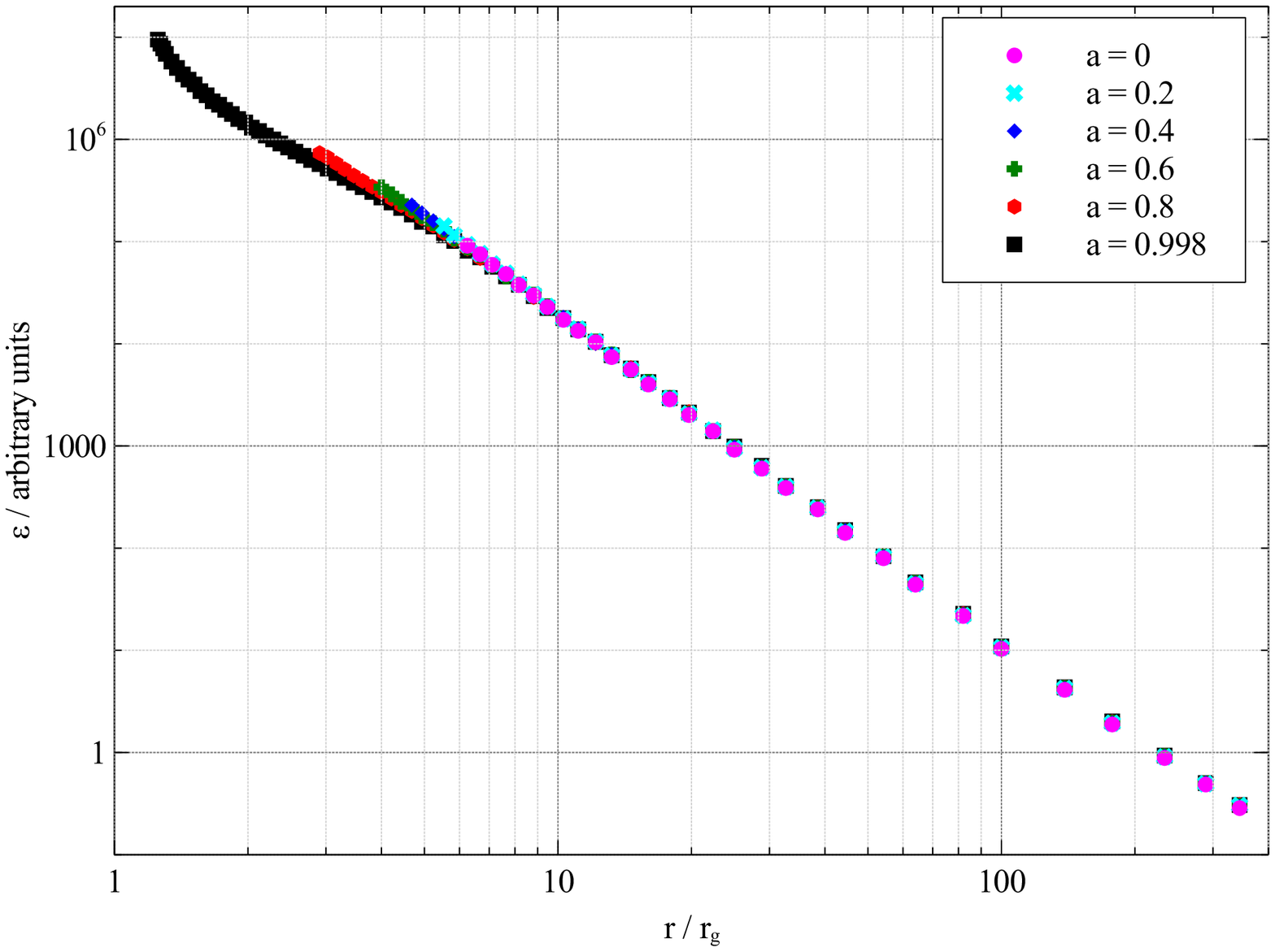}
\label{emis_spin.fig:axial}
}
\subfigure[]{
\includegraphics[width=80mm]{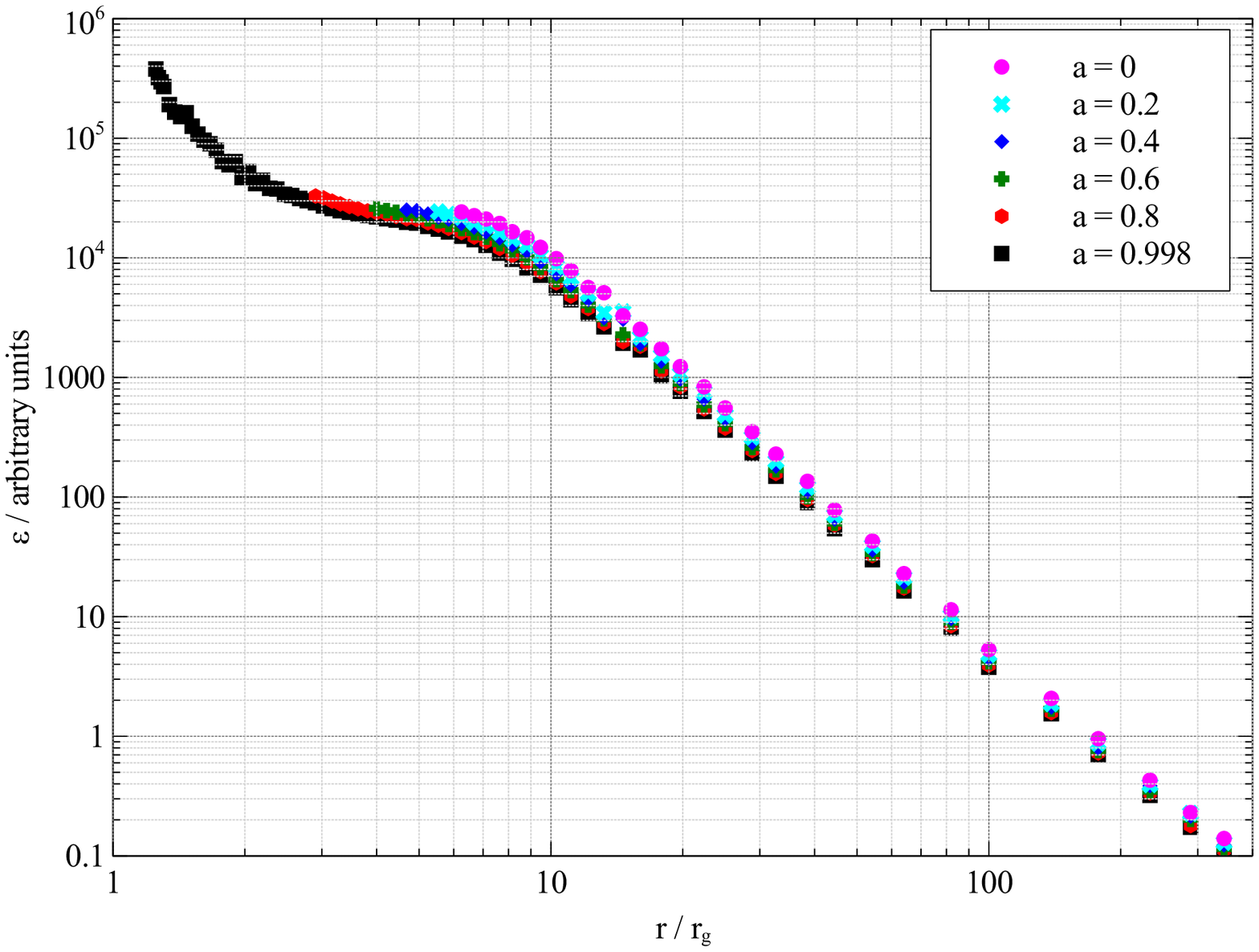}
\label{emis_spin.fig:ring}
}
\caption[]{Theoretical emissivity profiles of the accretion disc surrounding black holes with varying (dimensionless) spin parameter, $a$ for \subref{emis_spin.fig:axial} a stationary source located at a height 5\rg\ on the rotation axis and \subref{emis_spin.fig:ring} a ring source radius 3\rg\ located at a height 5\rg\ co-rotating with the disc below.}
\label{emis_spin.fig}
\end{minipage}
\end{figure*}

Fig. \ref{emis_spin.fig} shows theoretical emissivity profiles for the accretion disc around black holes with varying (dimensionless) spin parameter, $a$, for both an axial and orbiting ring sources. It can be seen that the spin parameter has little effect on the emissivity at specific locations upon the disc, with the only notable change in the profiles being the truncation of the disc at greater radii as the spin parameter decreases. As such, the steepening over the inner part of the accretion disc is only seen for rapidly spinning black holes, with no steepened inner part seen for $a\le 0.8$, though this is due to the lack of reflector at small radius where the spin is low rather than an intrinsic effect of the black hole spin. In the case of the source orbiting with the same angular velocity as the disc element below, reflection is slightly enhanced from the innermost regions for lower values of the spin owing to the angular velocity of a circular orbit being greater for smaller values of the spin parameter, increasing relativistic beaming.

This is, of course, only accounting for the propagation of X-rays around the black hole and assuming `idealised' reflection off of a razor thin accretion disc extending as close to the black hole as it is able to as defined by the innermost stable orbit. It may be that the accretion disc does not extend this far inwards, in the case of a truncated accretion disc with an advection-dominated hot accretion flow in the inner regions, though this solution would be more applicable to accretion discs around galactic black holes rather than to AGN. For example, \citet{beckwith+08} find in general relativistic magneto-hydrodynamic simulations that there is significant thermal dissipation within the innermost stable orbit, regardless of the spin of the black hole. The density of the accretion flow will drop greatly in this region (both due to the high temperatures and there being no stable circular orbit such that material plunges into the black hole), so little reflection, with which we are concerned here, will be observed from the accretion flow within the innermost stable orbit.

\subsection{Returning Radiation}
In general relativity, it is conceivable that X-rays reflected from one part of the accretion disc will be bent around the black hole back to the disc to be reflected a second time \citep{cunningham-76}. \citet{ross_fabian_ballantyne} considered the spectrum of the reflected radiation from an ionised slab (\textit{i.e.} an accretion disc) if a fraction of the reflected radiation is said to return to the reflector as incident radiation. They found that multiple reflection of returning radiation strengthens the broadened absorption and emission features such as the iron K line as well as steepening the softer part of the spectrum below 2\keV. We here simply consider the effects of gravitational light bending returning radiation to the disc in terms of the `raw' emissivity rather than considering the energy-dependent details of the reflection spectrum, since the emissivity profile is determined from the emission lines which are enhanced so we determine if this measurement is affected by returning radiation.

In order to compute the emissivity profile due to radiation returning to the accretion disc from the primary reflection component, the emission from the disc is modelled as the sum of isotropic point sources in (relativistic) Keplerian orbits in the disc at regular radial intervals. Due to the axisymmetry of the Kerr spacetime, each of these point sources represents emission (or more specifically reflection) from its respective annulus. The photon count from each source is scaled by the emissivity profile across the disc to account for the variation in incident flux and is multiplied by the (relativistic) area of the annulus.

This, of course, assumes that in the frame of the disc material, the reflected radiation is emitted isotropically while in reality anisotropic emission from atomic processes and Compton scattering will depend on the angle of incidence of the incoming radiation \citep{svoboda+10}, however provides a simple model which illustrates an upper limit of the effect of returning radiation.

Simulations reveal that as many as 50 per cent of the photons reflected from the accretion disc are returned to the disc (since for a typical X-ray source located above the plane of the accretion disc, the reflected flux will be concentrated on the innermost parts of the disc where the influence of the black hole focusing them back to the disc is the greatest).

\begin{figure}
	\centering
	\includegraphics[width=85mm]{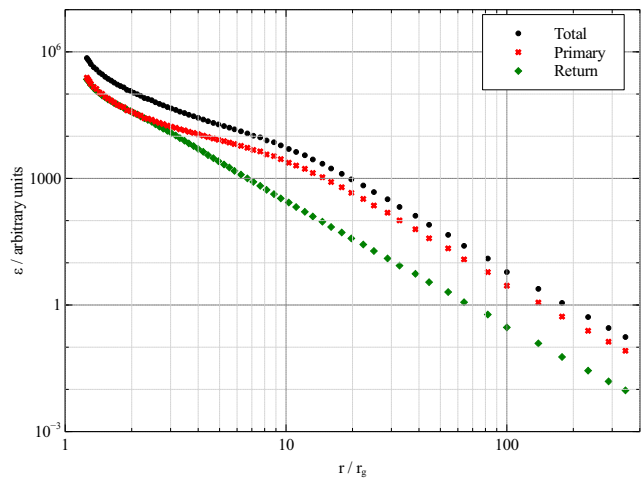}
	\caption{The effect of reflected radiation returning to the accretion disc to be reflected again on the overall emissivity profile. The emissivity profiles due to the primary X-ray source and returning X-rays are shown as red crosses and green diamonds, respectively, and the overall profile (the sum of these contributions) is shown as black points. Returning radiation has little effect on the overall emissivity profile.}
	\label{return.fig}
\end{figure}

Computing the returning radiation from a disc illuminated by a point source at a height 10\rg\ above the disc around a maximally spinning black hole yields an emissivity profile with an approximately constant power law slope, steepening over the innermost regions of the disc. Once the photon counts returning to the disc have been divided by the effective areas of the annuli, the emissivity due to returning radiation is an order of magnitude lower than that due to the primary X-ray source and with its almost constant power law slope, the returning radiation has little effect on the overall shape of the emissivity profile, as shown in Fig. \ref{return.fig}.

These results are consistent with those of \citet{agol_krolik} who show that for a maximally spinning black hole, up to 50 per cent of the radiation emitted from the accretion disc can be returned to the disc before finally being observed at infinity, though we here explicitly demonstrate that the distribution of re-reflected emission from the disc  means that returning radiation does not greatly affect the observed emissivity profile of the accretion disc. \citet{agol_krolik} also note that the fraction of the flux returned to the accretion disc for more slowly spinning black holes is less (as many of the returning rays end up within the innermost stable orbit in this case) so as returning radiation in the maximally spinning case does not affect the emissivity profile substantially, it will do so even less for more slowly spinning black holes. Furthermore, we can say that even including the effects of returning radiation, the black hole spin does not greatly affect the form of the emissivity profile save for the innermost extent of the accretion disc and thus the degree of steepening observed over the inner part of the disc.

\section{The case of 1H\,0707-495}
The emissivity profile of the accretion disc due to X-ray reflection in the narrow line Seyfert 1 galaxy, 1H\,0707-495 was determined by \citet{wilkins_fabian_2011a}, considering the relativistically broadened iron K line (whose rest-frame energy is 6.4\keV) to be composed of the sum of independent relativistically-blurred emission line components from successive annuli in the accretion disc and fitting for the relative contributions of these.

\begin{figure}
\centering
\subfigure[]{
\includegraphics[width=90mm]{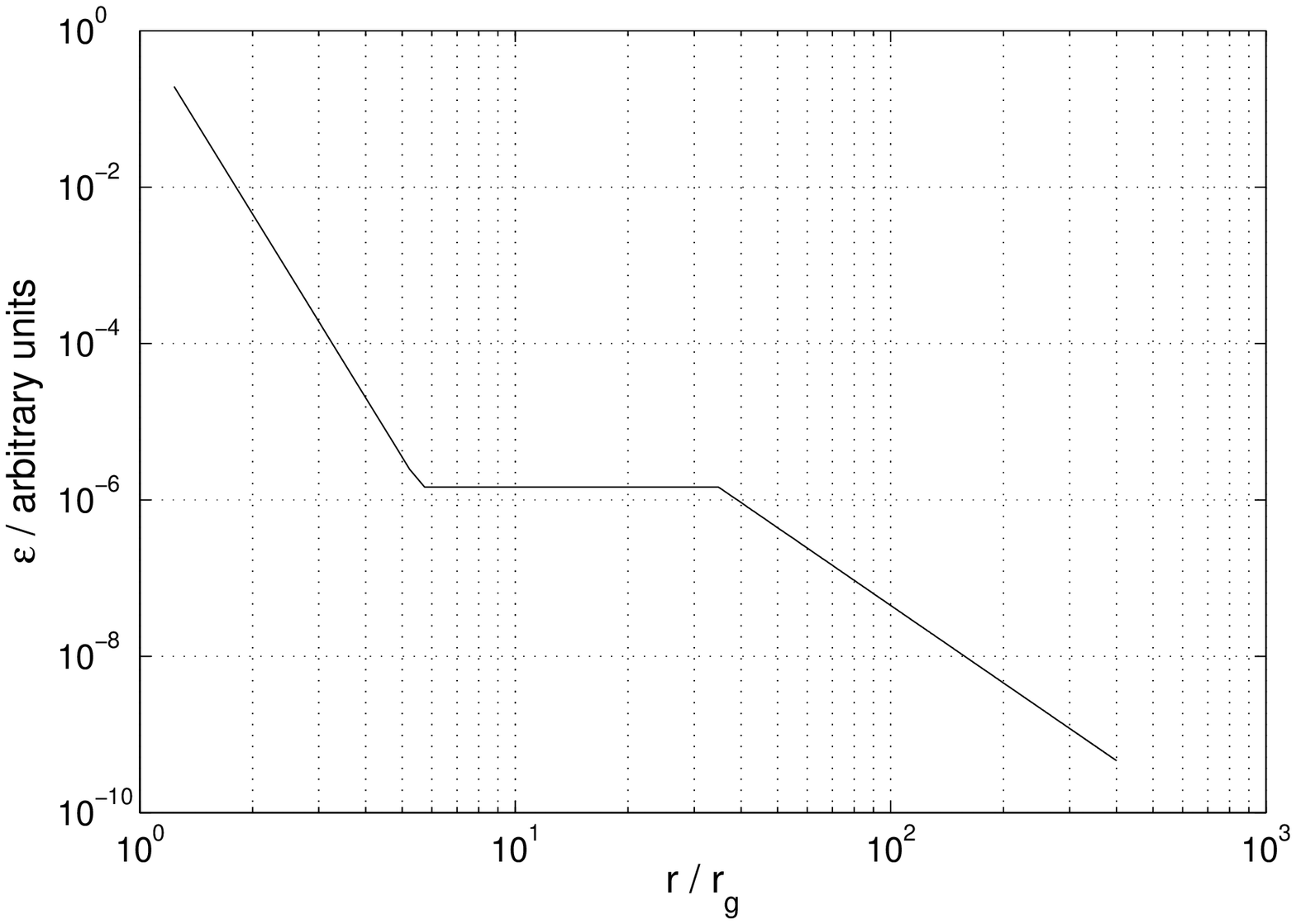}
\label{1h0707_jan08_emis.fig:obs}
}
\subfigure[]{
\includegraphics[width=90mm]{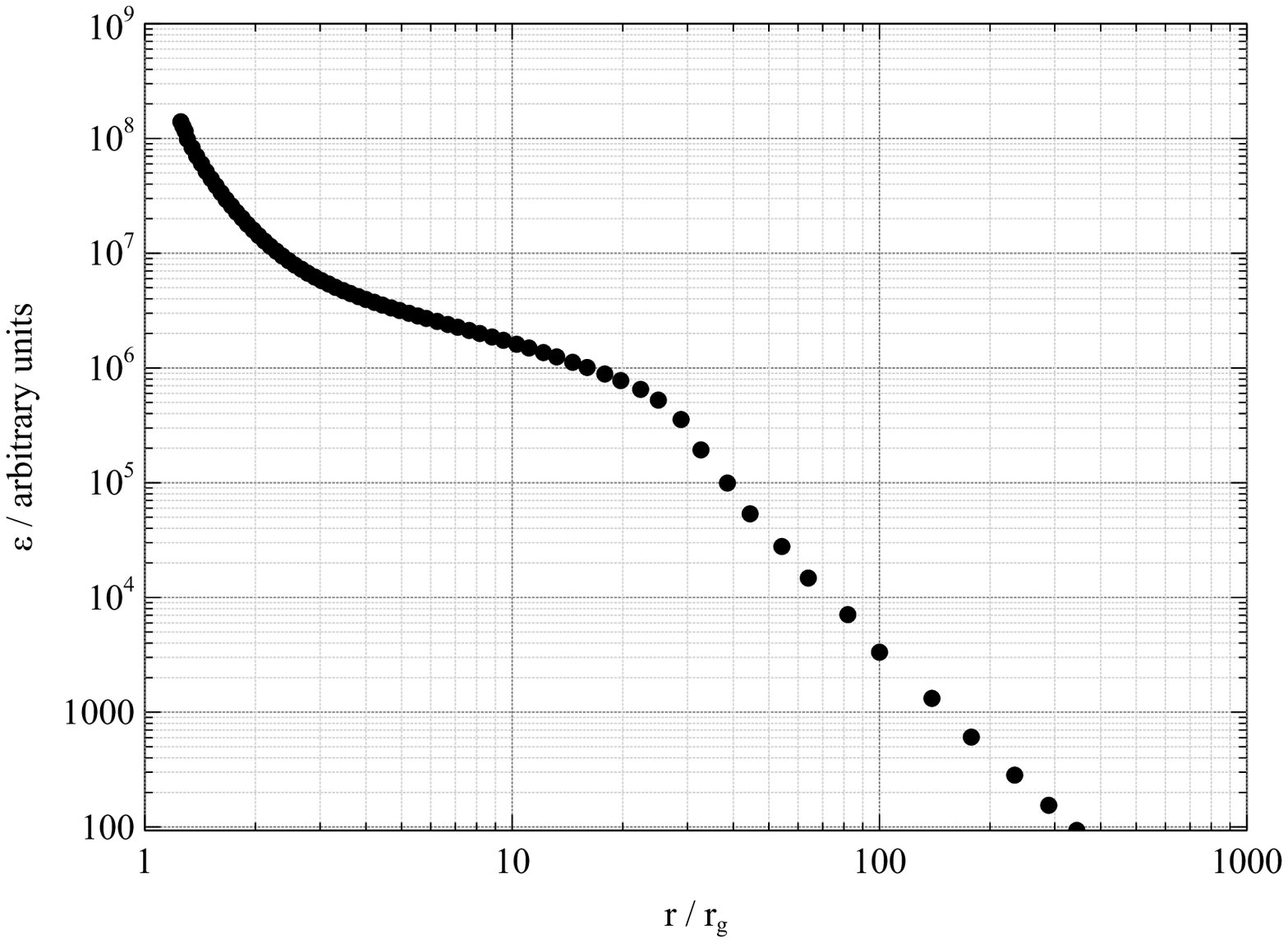}
\label{1h0707_jan08_emis.fig:theory}
}
\caption[]{ \subref{1h0707_jan08_emis.fig:obs} Emissivity profile determined for X-ray reflection from the accretion disc in 1H\,0707-495 in January 2008 by fitting for the relative contributions of components of the relativistically broadened iron K emission line from successive radii in the disc \citep{wilkins_fabian_2011a}, compared with \subref{1h0707_jan08_emis.fig:theory} a theoretical emissivity profile due to an extended X-ray source extending radially outwards to 30\rg\ and between 2 and 10\rg\ above the plane of the accretion disc. }
\label{1h0707_jan08_emis.fig}
\end{figure}

The accretion disc was found to have an emissivity profile approximated by a twice-broken power law, with a steep index of 7.8 over the innermost parts then flattening to almost constant emissivity (index zero) between 5.6\rg\ and 34.8\rg\ before tending to a constant index 3.3 over the outer parts of the disc (Fig. \ref{1h0707_jan08_emis.fig}).

\citet{fabian+09} and \citet{zoghbi+09} report reverberation time lags of around 30\s\ where variability in the reflection dominated component of the spectrum is seen to respond to that in the power law continuum after a time lag due to the finite light travel time from the primary X-ray source to the reflecting accretion disc. If the mass of the black hole is taken to be $3\times10^6\Msun$, as quoted in the literature \citep[see \textit{e.g.}][]{zhou_wang}, this implies that a significant portion of the observed X-rays originate from a source located within 2\rg\ of the reflector.

Comparing the observed emissivity profile to the theoretical emissivity profiles computed for a variety of source locations and geometries while applying the constraint that the source is as close to the disc as 2\rg\ implies that the source is located at a low height close to the disc plane while extending radially outwards to between 25 and 35\rg\ to account for the location of the outermost break point in the emissivity profile. A theoretical emissivity profile for a disc of emission between 2\rg\ and 10\rg\ above the disc, extending radially outward to 30\rg\ with constant luminosity throughout its extent is shown in Fig. \ref{1h0707_jan08_emis.fig}, where letting the source region extend to greater heights serves to flatten the middle section of the emissivity profile. To account for the 30\s\ reverberation signal, it is likely that the source will be more luminous in the lower regions closer to the disc, however this is also accounted for in the fact that a larger fraction of the primary emission will reach the disc from lower sources than those further from the disc, as can be seen in Fig. \ref{emis_ax.fig}, although a fully relativistic treatment of reflection time lags is required to quantify this fully and will be explored in future work.

The only notable discrepancy between theoretical models and the observed emissivity profile of 1H\,0707-495 is the extent of the inner region over which the profile is steepened, where fitting to the profile of the iron K emission line shows the power law emissivity profile to break from the steep inner index to the flat region around a radius of 5\rg, while the theoretical models produce emissivity profiles falling off steeply over the inner around 3\rg\ before the power law index gradually reduces to the flat region. In the observed spectra from which the emissivity profile of 1H\,0707-495 was determined, relatively little flux is in  the redshifted wing of the emission line since this originates from the innermost regions of the accretion disc closest to the black hole which subtend the smallest solid angle at the observer (not least because many of the rays reflected form this part of the disc will fall into the black hole). As such, the emissivity determined for this region is an upper limit, since to a certain degree, the emission from the inner parts of the disc can be increased with little effect seen by the observer as the excess photons fall into the black hole and are not observed. From the confidence limits on these contributions to the emission line, however, it is clear that some steepening of the emissivity profile is required to explain the broadened line profile in agreement with the theoretical calculations.

The spectral features contributed to the emission line by the inner regions of the disc out to 10\rg\ will appear in the energy range 3-5\keV\ (see. \textit{e.g.}. Fig. 1 of \citealt{wilkins_fabian_2011a}) over which the intrinsic energy resolution (FWHM) of the EPIC pn detector, used to obtain the spectra from which this emissivity profie was determined, is 150\eV. Spectral features distinguishing the emission from these inner annuli of the accretion disc will, to some degree, be smoothed out by the detector such that they will be harder to distinguish above the power law continuum.  The curving of the emissivity profile to the flattened middle section is therefore likely not fully resolved, rather we fit a broken power law to the observed emissivity profile.

The results of these ray tracing calculations combined with the observed emissivity profile of 1H\,0707-495 and the reflection time lags appear to suggest an emitting region surrounding the central black hole and covering the central parts of the accretion disc. This emitting region could consist a corona of hot electrons emitting X-rays through the inverse-Compton scattering of (thermal) seed photons emitted from the accretion disc. If this corona were optically thin to X-rays, as implied by the steep index of the power law continuum after Comptonisation of seed photons, for example a sparse plasma or even a region with emission originating from flares where magnetic reconnection takes place \citep{galeev+79,merloni_fabian,goyder_lasenby}, rays emitted from any part of it could reach the accretion disc, as modelled in ray tracing simulations here and the corona would not obscure the emission from the regions of the disc below.

\begin{figure}
\centering
\subfigure[]{
\includegraphics[width=90mm]{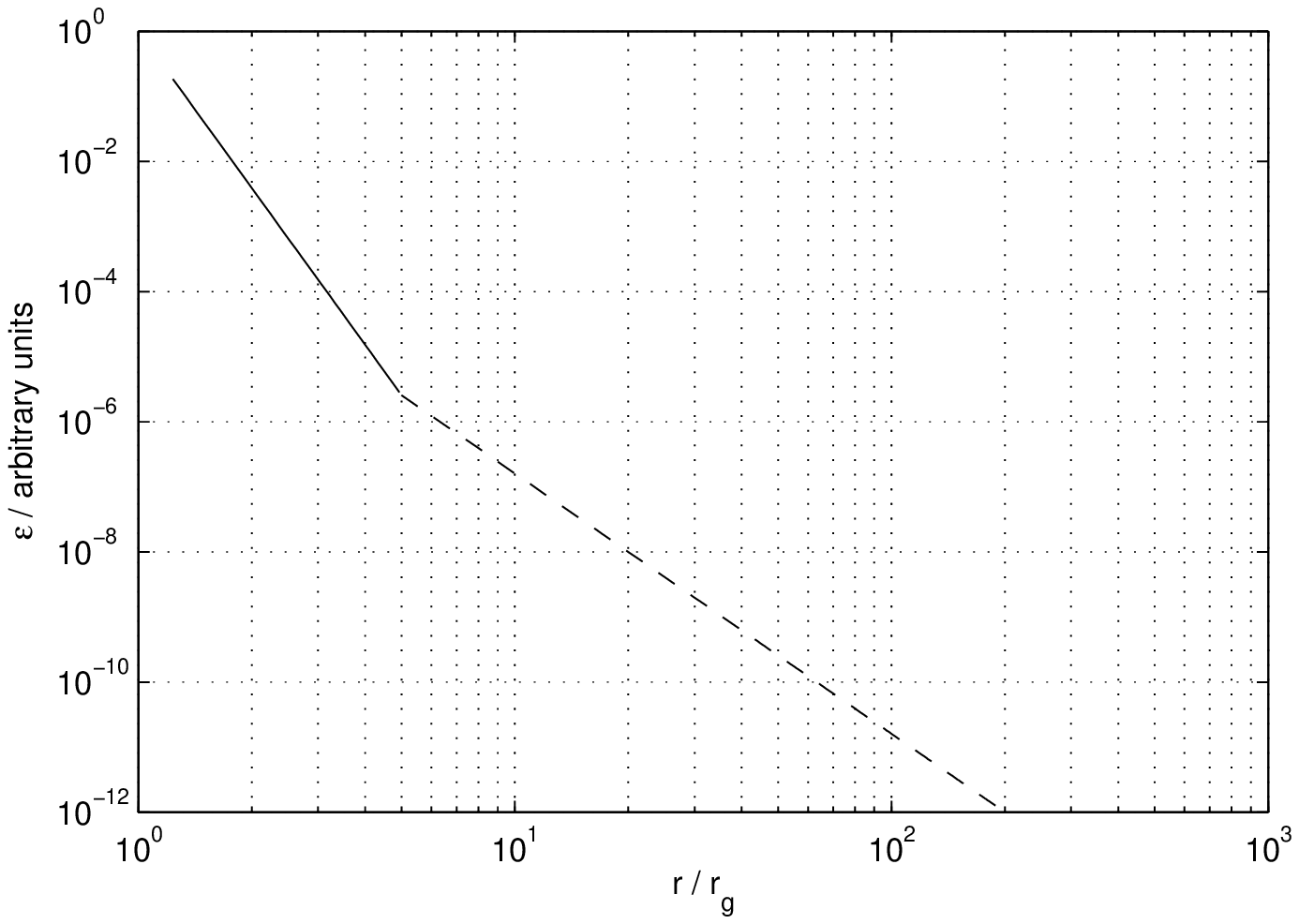}
\label{1h0707_jan11_emis.fig:obs}
}
\subfigure[]{
\includegraphics[width=90mm]{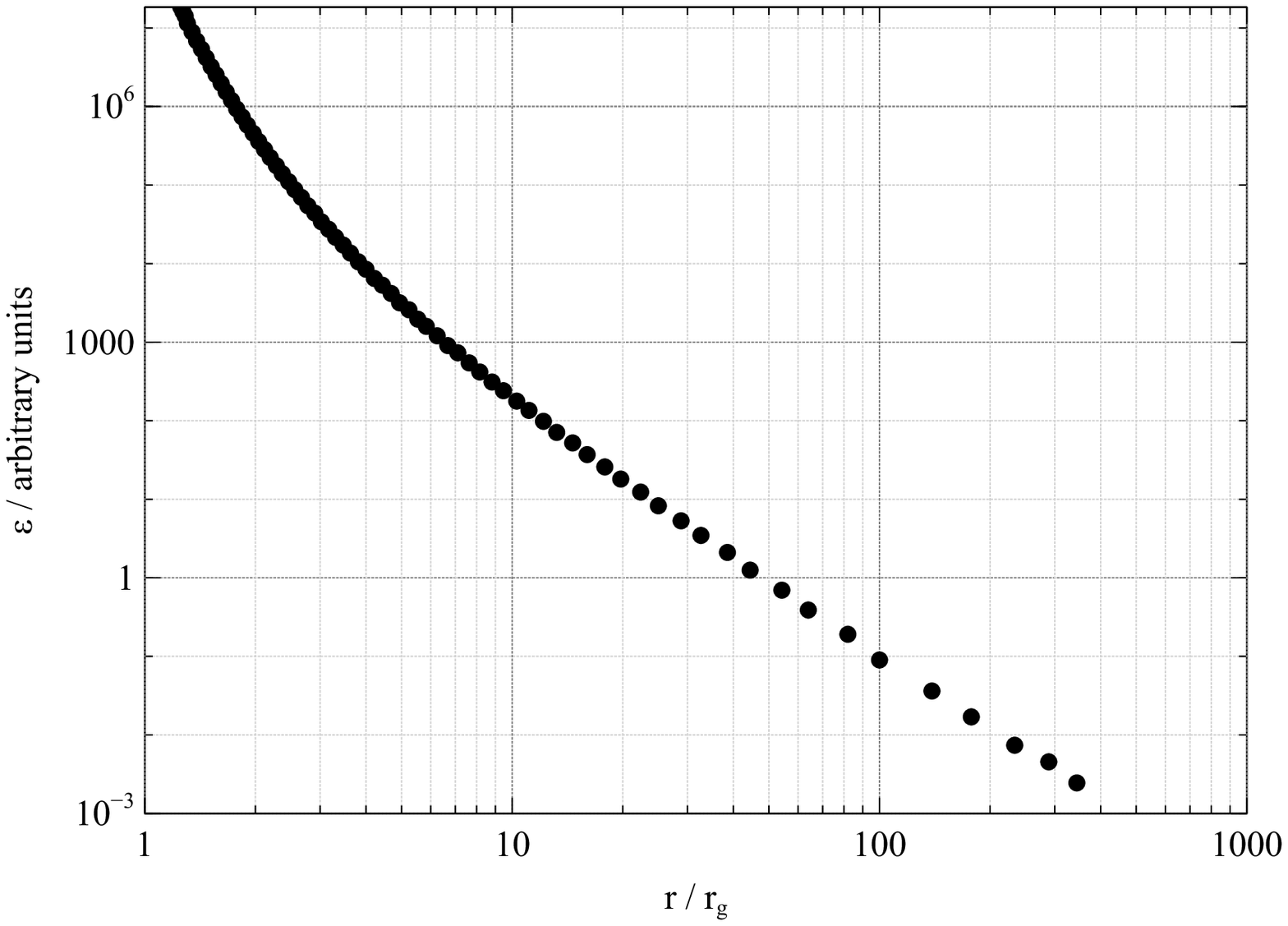}
\label{1h0707_jan11_emis.fig:theory}
}
\caption[]{ \subref{1h0707_jan11_emis.fig:obs} Emissivity profile determined for X-ray reflection from the accretion disc in 1H\,0707-495 in January 2011 by fitting for the relative contributions of components of the relativistically broadened iron K emission line from successive radii in the disc \citep{1h0707_jan11}, compared with \subref{1h0707_jan11_emis.fig:theory} a theoretical emissivity profile due to a compact X-ray source located on the X-ray source 1.5\rg\ above the plane of the accretion disc. }
\label{1h0707_jan11_emis.fig}
\end{figure}

In January 2011, 1H\,0707-495 dropped into a low flux state in which the observed spectrum was dominated by the reflection from the accretion disc, with little to no direct emission seen from the primary X-ray source \citep{1h0707_jan11}. The emissivity profile of the accretion disc was determined as for the January 2008 observation and was found to fall off steeply with a power law index of about 8 out to a radius of 5\rg\ before tending to a constant index slightly steeper than the classical value of 3 over the outer part of the disc (Fig. \ref{1h0707_jan11_emis.fig}). Comparing this to the theoretical emissivity profiles suggests that the X-ray source had collapsed down to a compact source around the rotation axis located at a low height of between 1.5 and 2\rg\ above the disc plane (Fig. \ref{1h0707_jan08_emis.fig}). Ray tracing simulations simply counting the number of photons that land on the accretion disc compared to the number able to escape to infinity (which will be seen as the power law continuum in the observed spectrum) demonstrate that such a compact source close to the black hole naturally explains the drop in the power law continuum emission with the majority of photons being focussed on to the accretion disc (and reflected) or into the black hole itself.

\section{Conclusions}
Numerical ray tracing simulations of the propagation of X-rays from a source in a corona surrounding the central black hole to the accretion disc have yielded theoretical predictions for the emissivity profile of the accretion disc due to X-ray reflection.

These calculations have demonstrated that the forms of the emissivity profiles observed through the profiles of relatively broadened emission lines from accretion discs in AGN (such as that from the narrow line Seyfert 1 galaxy 1H\,0707-495) arise naturally from the relativistic effects on the propagation of the rays around the black hole and on the accretion disc itself, with the simplest possible assumptions about the X-ray itself (assuming either an isotropic point source or an extended source of uniform intensity throughout its volume).

By comparing observed emissivity profiles to those computed theoretically for different locations and geometries of the source  along with consideration of the time lags between the continuum emission and reflection components in AGN, it is possible to constrain the location and extent of the primary X-ray source. In 1H\,0707-495, the emissivity profile obtained from the data appears to suggest an X-ray source as low as 2\rg\ above the plane of the accretion disc and extending outwards from the rotation axis to around 30\rg.

Considering the emissivity profile of the accretion disc in terms of the properties of the primary X-ray source has also allowed the change in state of 1H\'0707-495 observed in January 2011 (in which the spectrum was almost entirely reflection dominated with little direct continuum emission) to be interpreted in terms of the X-ray source collapsing to a confined region around the rotation axis, around 1.5-2\rg\ above the accretion disc.

\section*{Acknowledgements}
DRW thanks Jonathan Crass and Craig Mackay for access to the Tesla graphics processing card, provided to the Institute of Astronomy Instrumentation Group by the \textsc{nvidia} Corporation.

\bibliographystyle{mnras}
\bibliography{agn}

\appendix
\section{Constructing the Source Frame}
\label{source.app}
For completeness, we include here the derivation of the tetrad basis vectors defining the local rest frame of a source orbiting the rotation axis of the black hole in the Kerr spacetime (see also, \textit{e.g.}, \citealt{krolik+05, beckwith+08}).
When considering the dynamics of inertial frames and constructing tetrad basis vectors, it is often convenient to re-write the metric in the form (following \citealt{bhmath})
\[ ds^2 = e^{2\nu} c^2 dt^2 - e^{2\psi}(d\varphi - \omega dt)^2 - e^{2\mu_1} dr^2 - e^{2\mu_2} d\theta^2 \]
Where
\begin{eqnarray*}
	e^{2\nu} &=& \frac{\rho^2 \Delta}{\Sigma^2} \\
	e^{2\psi} &=& \frac{\Sigma^2 \sin^2\theta}{\rho^2} \\
	e^{2\mu_1} &=& \frac{\rho^2}{\Delta} \\
	e^{2\mu_2} &=& \rho^2 \\
	\omega &=& \frac{2\mu acr}{\Sigma^2}
\end{eqnarray*}
With
\[ \Sigma^2 = (r^2 + a^2)^2 - a^2\Delta\sin^2\theta \]
Hereafter, we shall set $\mu = c = 1$.

This definition of the metric co-efficients as exponentials allows their square roots, products and ratios to be easily written.

The source frame is defined by a tetrad of orthogonal basis vectors. By the equivalence principle, which states that in a local, freely-falling laboratory the laws of physics reduce to those of special relativity, the local spacetime in the source frame is flat and the tetrad basis vectors satisfy
\begin{eqnarray}
	\mathbf{e}'_{(a)} \cdot \mathbf{e}'_{(b)} &=& \eta_{(a)(b)} \\
	\label{orthcondition.equ}	
	g_{\mu\nu} e_{(a)}^\mu e_{(b)}^\nu &=& \eta_{(a)(b)}
\end{eqnarray}
Note that bracketed subscripts indicate directions in the tetrad basis and those not bracketed directions in the co-ordinate basis. $\eta_{(a)(b)} = \mathrm{diag}(1,-1,-1,-1)$ is the Minkowski (flat) space metric.

The timelike tetrad basis vector, $\mathbf{e}'_{(t)}$ is, by definition of the instantaneous rest frame, parallel to the source's 4-velocity, $\mathbf{u}_0$ (since in its rest frame, the 4-velocity must have no spacelike component as its 3-velocity is zero). The 4-velocity of the emitting particle orbiting at angular velocity $\frac{d\varphi}{dt} = \Omega$ is
\begin{eqnarray*}
	\mathbf{e}'_{(t)} &=& \mathbf{u} \quad = (u^t, \dot{\overrightarrow{r}}) = \left(u^t, \frac{d\overrightarrow{r}}{dt}\dot{t}\right) = u^t\left(1, \frac{d\overrightarrow{r}}{dt}\right) \\
	\mathbf{e}'_{(t)} &=& u^t(1,0,0,\Omega)
\end{eqnarray*}
The time-like component, $u^t$, is found from the condition for the motion of a massive particle $g_{\mu\nu}u^\mu u^\nu = c^2$ (again taking $c=1$).
\[	g_{tt}u^t u^t + 2g_{t\varphi} u^t u^\varphi + g_{\varphi\varphi} u^\varphi u^\varphi = 1 \]
\[ (e^{2\nu} - \omega^2e^{2\psi})(u^t)^2 + 2\omega e^{2\psi}\Omega (u^t)^2 - e^{2\psi}\Omega^2(u^t)^2 = 1 \]
\[ u^t = \frac{e^{-\nu}}{\sqrt{1-e^{2(\psi-\nu)}(\Omega - \omega)^2}} \]

Since the 4-velocity has no component in either the $r$ or $\theta$ directions and the Kerr metric is diagonal in these directions, two of the orthogonal space-like basis vectors are
\begin{eqnarray*}
	\mathbf{e}'_{(3)} &=& \left(0,\frac{\sqrt{\Delta}}{\rho},0,0\right) \\
	\mathbf{e}'_{(2)} &=& \left(0,0,\frac{1}{\rho},0\right)
\end{eqnarray*}
Finally, the fourth orthogonal tetrad basis vector must take the form
\[ \mathbf{e}'_{(1)} = \left( e_{(1)}^t, 0, 0, e_{(1)}^\varphi \right) \]
and satisfy the conditions (\ref{orthcondition.equ})
\begin{eqnarray}
	\label{rote1cond1.equ}
	\mathbf{e}'_{(t)} \cdot \mathbf{e}'_{(1)} &=& 0 \\
	\label{rote1cond2.equ}
	\mathbf{e}'_{(1)} \cdot \mathbf{e}'_{(1)} &=& -1
\end{eqnarray}
Condition (\ref{rote1cond1.equ}) gives
\[ g_{tt} e_{(t)}^t e_{(1)}^t + g_{t\varphi} e_{(t)}^t e_{(1)}^\varphi + g_{\varphi t} e_{(t)}^\varphi e_{(1)}^t + g_{\varphi\varphi} e_{(t)}^\varphi e_{(1)}^\varphi = 0 \]
\begin{equation*}
	\label{rote1phi.equ}
	e_{(1)}^\varphi = \frac{e^{2\nu} - \omega^2e^{2\psi} + \Omega\omega e^{2\psi}}{e^{2\psi}(\Omega-\omega)}e_{(1)}^t
\end{equation*}
And from (\ref{rote1cond2.equ}),
\begin{equation*}
g_{tt} e_{(1)}^t e_{(1)}^t + 2g_{t\varphi} e_{(1)}^t e_{(1)}^\varphi + g_{\varphi\varphi} e_{(1)}^\varphi e_{(1)}^\varphi = -1
\end{equation*}
Substituting using (\ref{rote1phi.equ}) and solving gives
\begin{eqnarray*}
	e_{(1)}^t &=& \frac{e^{\psi-\nu}(\Omega - \omega)}{\sqrt{e^{2\nu} - e^{2\psi}(\Omega - \omega)^2}} \\
	e_{(1)}^\varphi &=& \frac{e^{-\nu-\psi}(e^{2\nu} + \Omega\omega e^{2\psi} - \omega^2e^{2\psi})}{\sqrt{e^{2\nu} - e^{2\psi}(\Omega - \omega)^2}}
\end{eqnarray*}
Where the signs have been selected to form a right-handed basis.

Thus, the tetrad basis vectors in the frame of a massive observer orbiting a Kerr black hole at angular velocity $\frac{d\varphi}{dt} = \Omega$ are
\begin{eqnarray*}
	\label{et.equ}	
	\mathbf{e}'_{(t)} &=& \left( \frac{e^{-\nu}}{\sqrt{1-e^{2(\psi-\nu)}(\Omega - \omega)^2}}, 0, 0, \right.\\
		&\ &\qquad\qquad \left. \frac{e^{-\nu}\Omega}{\sqrt{1-e^{2(\psi-\nu)}(\Omega - \omega)^2}} \right) \\
	\label{e1.equ}
	\mathbf{e}'_{(1)} &=& \left( \frac{e^{\psi-\nu}(\Omega - \omega)}{\sqrt{e^{2\nu} - e^{2\psi}(\Omega - \omega)^2}}, 0, 0, \right. \\
		&\ & \qquad\qquad \left. \frac{e^{-\nu-\psi}(e^{2\nu} + \Omega\omega e^{2\psi} - \omega^2e^{2\psi})}{\sqrt{e^{2\nu} - e^{2\psi}(\Omega - \omega)^2}} \right) \\
	\label{e2.equ}
	\mathbf{e}'_{(2)} &=& \left(0,0,\frac{1}{\rho},0\right) \\
	\label{e3.equ}
	\mathbf{e}'_{(3)} &=& \left(0,\frac{\sqrt{\Delta}}{\rho},0,0\right)
\end{eqnarray*}
Taking $\mathbf{e}'_{(3)}$, $\mathbf{e}'_{(2)}$ and $\mathbf{e}'_{(1)}$ as the $z$ (polar), $x$ and $y$ axes respectively (forming a right-handed co-ordinate system), the 4-momentum of a photon energy $E_0$ emitted at declination $\alpha$ and azimuth $\beta$ in the emitter's frame is written
\[ \mathbf{p}' = \left(E_0, E_0\sin\alpha\sin\beta, E_0\sin\alpha\cos\beta, E_0\cos\alpha\right) \]
And transforming back into the Boyer-Lindquist co-ordinate basis,
\begin{eqnarray*}
	\dot{t} &=& p^{(t)}e_{(t)}^t + p^{(1)}e_{(1)}^t \\
	\dot{r} &=& p^{(3)}e_{(3)}^r \\
	\dot{\theta} &=& p^{(2)}e_{(2)}^\theta \\
	\dot{\varphi} &=& p^{(t)}e_{(t)}^\varphi + p^{(1)}e_{(1)}^\varphi
\end{eqnarray*}

\label{lastpage}

\end{document}

%% file: defn.tex
\newcommand{\s}{\rm\thinspace s}

\newcommand{\Msun}{\hbox{$\rm\thinspace M_{\odot}$}}

\newcommand{\keV}{\rm\thinspace keV}
\newcommand{\eV}{\rm\thinspace eV}

\newcommand{\rg}{\rm\thinspace $r_\mathrm{g}$}

